\documentclass[twocolumn,showpacs,preprintnumbers,amsmath,amssymb,prb]{revtex4}

\usepackage{graphicx}
\usepackage{dcolumn}
\usepackage{bm}

\begin{document}

\title{Transfer-matrix approach to multiband Josephson junctions}

\author{S.~Graser and T.~Dahm}

\affiliation{Institut f\"ur Theoretische Physik, Universit\"at T\"ubingen,\\
             Auf der Morgenstelle 14, D-72076 T\"ubingen, Germany}

\date{\today}

\begin{abstract}
We study the influence of multiple bands on the properties of Josephson junctions. In particular we focus on the two gap superconductor magnesium diboride. We present a formalism to describe tunneling at a point contact between two MgB$_2$ electrodes generalizing the transfer-matrix approach to multiple bands. A simple model is presented to determine the effective hopping amplitudes between the different energy bands as a function of the misorientation angle of the electrodes. We calculate the critical current and the current-voltage characteristics for N-I-S and S-I-S contacts with different orientation for junctions with both high and low transparency. We find that interband tunneling processes become increasingly important with increasing misorientation angle. This is reflected in certain features in the differential tunneling conductance in both the tunneling limit as well as for multiple Andreev reflections.
\end{abstract}


\maketitle

\section{\label{sec:intro} Introduction}
The two gap superconductor MgB$_2$ is considered to be a good candidate for superconducting devices due to its comparatively high critical temperature, its easy handling and cheap preparation. Presently, its use in Josephson junctions is being investigated~\cite{shimakage,ueda,chen,singh,vanzalk}. One of its peculiar properties is the presence of two distinct superconducting gaps existing on different bands at the Fermi surface. These bands are $\pi$ and $\sigma$ bands arising from the Boron $p_z$ and $p_x$/$p_y$ orbitals, respectively \cite{kortus, dahmreview}. These two types of bands possess different parity with respect to reflection at the Boron plane. This different parity suppresses scattering and transitions between the two types of bands, which is thought to be the main reason for the exceptional stability of the two gaps against
impurity scattering \cite{mazin}.

In the vicinity of a Josephson junction this parity can be broken, if the two MgB$_2$ electrodes are grown with a misorientation angle of the crystal $c$-axis directions on both sides of the junction. The size of the misorientation angle would provide a means of tuning the strength of interband transitions between the two sides of the junction. This is what we wish to explore in the present work. In particular, we want to study the differential conductance of such a junction and demonstrate how these interband processes show up in the tunneling conductance and in Andreev reflections.

Due to the importance of Josephson junctions in numerous applications the Josephson effect has been subject of intense theoretical study. The current-phase relation as well as the oscillation of the electron transport in a voltage biased point contact known as the ac-Josephson effect has been theoretically investigated for different situations~\cite{gunsenheimer,averin,cuevas1} including unconventional superconductivity~\cite{cuevas2}, spin active magnetic barriers~\cite{fogelstrom,zhao1,zhao2} and the effect of pair breaking due to disorder~\cite{zaitsev1}. It has been shown that within the quasiclassical theory interfaces can be described by effective boundary conditions~\cite{zaitsev2} that can be included by mainly two different approaches: the scattering-matrix approach and the transfer-matrix approach~\cite{cuevas2}. Both are formally equivalent but the range of applicability is different and the problem under consideration can help to decide which one should be chosen. Here, we are going to generalize the transfer-matrix approach to the case of a multiband Josephson junction and use it to calculate the differential conductance of a MgB$_2$ Josephson junction with misoriented electrodes. The effects of quasiparticle and Josephson tunneling between two multiband electrodes with application to MgB$_2$ have been theoretically discussed by Brinkman et al. within an extended BTK-model~\cite{brinkman} explaining convincingly the absence of a double gap structure in $c$-axis tunneling spectra. Novel Josephson effects between multiband and singleband superconductors including the claim of a phase shift of $\pi$ between the different gaps on the multiband side have been theoretically discussed by Agterberg et al.~\cite{agterberg} showing the rich and interesting physical content of Josephson effects in multiband systems. The experimental observation of subharmonic gap structures in MgB$_2$ single crystal junctions~\cite{martinez} and MgB$_2$/Nb micro-constrictions~\cite{giubileo} due to multiple Andreev reflections can be understood within a multiband transfer-matrix approach. We will start in the next section with a description of the transfer-matrix model generalizing the approach to include the effect of multiple gaps. In the third section we will derive the current expression for both the equilibrium case without an applied voltage and the non-equilibrium case with an applied voltage. In the fourth section we will model the effective hopping amplitudes at the interface making use of some general considerations. In the fifth section we will show the results of our calculations while in the last section we will conclude.

\section{\label{sec:tmapproach} The transfer-matrix approach for multiple bands}
The quasiclassical theory of superconductivity has been proven to be a powerful tool to consider spatially inhomogeneous problems in equilibrium and non-equilibrium, for example to calculate the local quasiparticle density of states around vortices or in the vicinity of boundaries. However it is only valid within the quasiclassical limit  $k_F \xi \gg 1$ and can therefore only describe situations with slowly varying fields and potentials (on the scale of the coherence length). The interface between two superconducting regions of different orientation or between a normal metal and a superconductor represents a strong pertubation on a lengthscale much smaller than the coherence length which is -- in principle -- out of the range of validity of the quasiclassical theory. To describe this strong pertubation within the quasiclassical limit one has to find effective boundary conditions that connect the solutions on both sides. The first formulation of these boundary conditions for nonmagnetic interfaces has been found by Zaitsev~\cite{zaitsev2}. Afterwards they have been generalized by Millis et al.~\cite{millis} for magnetic active interfaces and have been explicitly solved by Shelankov~\cite{shelankov} for equilibrium problems and complemented by Eschrig~\cite{eschrig} for non-equilibrium problems within the powerful Riccati parametrization of the quasiclassical theory~\cite{schopohl}. To describe the complicated processes at the interface between two superconductors including multiple Andreev scattering the transfer-matrix approach has been proven to be as well suitable as the scattering-matrix approach but gives a more intuitive understanding. To describe what is happening at the boundary between two superconductors with several Fermi surfaces we will generalize in this work the transfer-matrix description of Cuevas and Fogelstr\"om~\cite{cuevas2}  and Kopu et al.~\cite{kopu} for multiple Fermi surfaces. The basic idea of this approach is to consider two decoupled superconducting electrodes described by two electrode Green's functions $\check{g}_{\infty,L/R}$ that are calculated in a left ($L$) and a right ($R$) half-space with an inpenetrable surface. In a second step we allow virtual hopping processes between the two electrodes including the possibility of intraband and interband hopping. This serial expansion is exactly summed up to infinite order using the technique known from the Dyson formalism. As has been shown by Kopu et al. this pertubative transfer-matrix approach is equivalent to other methods and can be used to calculate interfaces with arbitrary transparency. In the following we will repeat the steps described by Cuevas and Fogelstr\"om and we will discuss the consequences that arise from the multiband character of the superconducting pairing interaction. To describe the hopping processes we start with a phenomenological transfer Hamiltonian
\begin{equation}
\hat {H}_T = \sum_\sigma \sum_{\alpha,\alpha'} \left( \hat{c}^\alpha_{L,\sigma} \right)^\dagger v_{LR}^{\alpha \alpha'}  \hat{c}^{\alpha'}_{R,\sigma} + \left( \hat{c}^\alpha_{R,\sigma} \right)^\dagger v_{RL}^{\alpha \alpha'}  \hat{c}^{\alpha'}_{L,\sigma}
\end{equation}
where $\alpha$ and $\alpha'$ denote band indices corresponding to a special Fermi surface and introducing the effective hopping amplitudes $v_{LR}^{\alpha \alpha'}$ describing hopping processes from band $\alpha'$ in the right electrode to band $\alpha$ in the left electrode, and $v_{RL}^{\alpha \alpha'}$ correspondingly. In this formulation we have enlarged our Hilbert space to include not only the band index but also the quantum number of the electrodes. The full Green's function and the $T$-matrix can be written as 
\begin{equation}
\tilde{\check{\mathcal{G}}} = \left( \begin{array}{cc}
\check{\mathcal{G}}_{LL} & \check{\mathcal{G}}_{LR} \\
\check{\mathcal{G}}_{RL} & \check{\mathcal{G}}_{RR}
\end{array} \right),
\hspace{1em}
\tilde{\check{\mathcal{T}}} = \left( \begin{array}{cc}
\check{\mathcal{T}}_{LL} & \check{\mathcal{T}}_{LR} \\
\check{\mathcal{T}}_{RL} & \check{\mathcal{T}}_{RR}
\end{array} \right)
\end{equation}
taking into account reflection ($LL$ and $RR$) and transmission processes ($LR$ and $RL$). The Green's functions $\check{\mathcal{G}}_{ij}$ and $T$-matrices $\check{\mathcal{T}}_{ij}$ themselves are $N \times N$-matrices in the enlarged band space where $N$ is the number of bands
\begin{equation}
\check{\mathcal{G}}_{ij} = \left( \begin{array}{ccc}
\check{G}_{ij}^{11} & \cdots & \check{G}_{ij}^{1N} \\
\vdots & \ddots & \vdots \\
\check{G}_{ij}^{N1} & \cdots &  \check{G}_{ij}^{NN}
\end{array} \right),
\check{\mathcal{T}}_{ij} = \left( \begin{array}{ccc}
\check{T}_{ij}^{11} & \cdots & \check{T}_{ij}^{1N} \\
\vdots & \ddots & \vdots \\
\check{T}_{ij}^{N1} & \cdots &  \check{T}_{ij}^{NN}
\end{array} \right)
\end{equation}
consisting of 2$\times$2-matrices in Keldysh space, denoted by the check symbol
\begin{equation}
\check{G}_{ij}^{\alpha \alpha'} = \left( \begin{array}{cc}
\hat{G}_{ij}^{R,\alpha \alpha'} & \hat{G}_{ij}^{K,\alpha \alpha'} \\
\hat{0} & \hat{G}_{ij}^{A,\alpha \alpha'} 
\end{array} \right)
\end{equation}
The retarded, advanced and Keldysh Green's functions are 2$\times$2 Nambu-Gor'kov matrices in particle-hole space while we will neglect the spin degrees of freedom since we are concerned with spin singlet systems and we assume only non-magnetic interactions. Besides the full Green's functions we also have to consider the Green's functions of the unperturbed left and right electrodes, which take a diagonal form in the space of the two electrode quantum numbers, and we can write down the coupling parameters as off-diagonal elements in the same notation: 
\begin{equation}
\tilde{\check{\mathcal{G}}}_\infty  = \left( \begin{array}{cc}
\check{\mathcal{G}}_{\infty,L} & 0 \\
0 & \check{\mathcal{G}}_{\infty,R}
\end{array} \right),
\hspace{1em}
\tilde{\check{v}} = \left( \begin{array}{cc}
0 & \check{v}_{LR} \\
\check{v}_{RL} & 0
\end{array} \right)
\end{equation}
Since we want to neglect direct interband scattering in the decoupled left and right electrodes the self-energy and therefore also the unpertubed Green's functions take diagonal form in band space
\begin{equation}
\check{\mathcal{G}}_{\infty,i} = \left( \begin{array}{ccc}
\check{G}_{\infty,i}^{1} & \cdots & 0 \\
\vdots & \ddots & \vdots \\
0 & \cdots &  \check{G}_{\infty,i}^{N}
\end{array} \right),
\check{v}_{ij} = \left( \begin{array}{ccc}
\check{v}_{ij}^{11} & \cdots & \check{v}_{ij}^{1N} \\
\vdots & \ddots & \vdots \\
\check{v}_{ij}^{N1} & \cdots &  \check{v}_{ij}^{NN}
\end{array} \right)
\end{equation}
We will see in the following that the full matrix structure of the $\check{v}_{ij}$ in band space prevents us from an easy decoupling of the different matrix elements as it can be done in the space of the electrode quantum numbers. Therefore we have to solve the full matrix problem in band space. With regard to the application to the two-band superconductor MgB$_2$, where interband scattering processes are suppressed~\cite{mazin}, we will go on with these simplified diagonal electrode Green's functions but of course it is straight forward to extend the considerations to a full matrix problem including interband scattering. The transmission processes can now be summed up in form of a $T$-matrix equation with 
\begin{equation}
\tilde{\check{\mathcal{T}}} = \tilde{\check{v}}  + \tilde{\check{v}}  \circ \tilde{\check{\mathcal{G}}}_\infty \circ \tilde{\check{\mathcal{T}}} = \tilde{\check{v}}  + \tilde{\check{v}}  \circ \tilde{\check{\mathcal{G}}} \circ \tilde{\check{v}} 
\label{eq:t-matrix}
\end{equation}
This equation has to be complemented by the Dyson equation which reads 
\begin{equation}
\tilde{\check{\mathcal{G}}} = \tilde{\check{\mathcal{G}}}_\infty + \tilde{\check{\mathcal{G}}}_\infty \circ \tilde{\check{\mathcal{T}}} \circ \tilde{\check{\mathcal{G}}}_\infty = \tilde{\check{\mathcal{G}}}_\infty + \tilde{\check{\mathcal{G}}}_\infty \circ \tilde{\check{v}} \circ \tilde{\check{\mathcal{G}}} 
\end{equation}
Using the first identity of Eq.~(\ref{eq:t-matrix}) repeatedly we can write 
\begin{equation}
\tilde{\check{\mathcal{T}}} = \tilde{\check{v}}  + \tilde{\check{v}}  \circ \tilde{\check{\mathcal{G}}}_\infty \circ  \tilde{\check{v}}  + \tilde{\check{v}}  \circ \tilde{\check{\mathcal{G}}}_\infty \circ  \tilde{\check{v}}  \circ \tilde{\check{\mathcal{G}}}_\infty \circ \tilde{\check{\mathcal{T}}} 
\end{equation}
If we now employ the diagonal structure of the unperturbed Green's functions und the off-diagonal structure of the hopping matrices in the space of the electrode quantum numbers we get the following set of decoupled equations for the components of  $\tilde{\check{\mathcal{T}}}$. We note the diagonal (reflection) elements of the $T$-matrix as 
\begin{eqnarray}
\check{\mathcal{T}}_{LL}  & = &  \check{v}_{LR}  \circ \check{\mathcal{G}}_{\infty,R} \circ \check{v}_{RL}  \nonumber \\
& & + \check{v}_{LR}  \circ \check{\mathcal{G}}_{\infty,R} \circ \check{v}_{RL} \circ  \check{\mathcal{G}}_{\infty,L} \circ \check{\mathcal{T}}_{LL}  \\
\check{\mathcal{T}}_{RR}  & = &  \check{v}_{RL}  \circ \check{\mathcal{G}}_{\infty,L} \circ \check{v}_{LR}  \nonumber \\
& & + \check{v}_{RL}  \circ \check{\mathcal{G}}_{\infty,L} \circ \check{v}_{LR} \circ  \check{\mathcal{G}}_{\infty,R} \circ \check{\mathcal{T}}_{RR}  
\end{eqnarray}
and we can write the off-diagonal (transmission) elements as
\begin{eqnarray}
\check{\mathcal{T}}_{LR}  & = &  \check{v}_{LR}  + \check{v}_{LR}  \circ \check{\mathcal{G}}_{\infty,R} \circ \check{v}_{RL} \circ  \check{\mathcal{G}}_{\infty,L} \circ \check{\mathcal{T}}_{LR}  \\
\check{\mathcal{T}}_{RL}  & = &  \check{v}_{RL}  + \check{v}_{RL}  \circ \check{\mathcal{G}}_{\infty,L} \circ \check{v}_{LR} \circ  \check{\mathcal{G}}_{\infty,R} \circ \check{\mathcal{T}}_{RL}  
\end{eqnarray}
Now we can perform the quasiclassical integration since there are no functions with spatial arguments from both sides of the interface any more. The quasiclassical $\xi$-integration reads 
\begin{equation}
\check{g}_{\infty,i} (\hat{p}_F,t,t') = \frac{1}{\pi} \int d \xi \check{\tau}_3 \sqrt{\check{\rho}_F}^{-1} \check{\mathcal{G}}_{\infty,i} (\vec{p}_F,t,t') \sqrt{\check{\rho}_F}^{-1}
\end{equation}
where the matrix $\sqrt{\check{\rho}_F}$ gives the square root of the density of states at the Fermi level in the normal state. This matrix is proportional to the identity matrix in Keldysh space but has different elements on the diagonal in band space 
\begin{equation}
\sqrt{\rho_F} = \left( \begin{array}{ccc}
\sqrt{N_F^{1}} & \cdots & 0 \\
\vdots & \ddots & \vdots \\
0 & \cdots &  \sqrt{N_F^{N}}
\end{array} \right)
\end{equation}
We can go on defining the quasiclassical transfer and hopping matrices as 
\begin{eqnarray}
\check{v}_{ij} (\hat{p}_F,\hat{p}_F') & = & \pi \sqrt{\check{\rho}_F} \check{v}_{ij} (\vec{p}_F,\vec{p}_F') \sqrt{\check{\rho}_F} \check{\tau}_3, \\
\check{t}_{ij} (\hat{p}_F,\hat{p}_F',t,t') & = & \pi \sqrt{\check{\rho}_F}  \check{\mathcal{T}}_{ij} (\vec{p}_F,\vec{p}_F',t,t') \sqrt{\check{\rho}_F}  \check{\tau}_3
\end{eqnarray} 
Finally we can write down a set of equations defining the quasiclassical transfer matrices 
\begin{eqnarray}
\check{t}_{LL}  & = & \left\langle \check{v}_{LR}  \otimes \check{g}_{\infty,R} \otimes \check{v}_{RL} \right\rangle_{\hat{p}_F}  \nonumber \\
& & + \left\langle \check{v}_{LR}  \otimes \check{g}_{\infty,R} \otimes \check{v}_{RL} \otimes  \check{g}_{\infty,L} \otimes \check{t}_{LL} \right\rangle_{\hat{p}_F} \label{eq:tLL} \\
\check{t}_{RR}  & = &  \left\langle \check{v}_{RL}  \otimes \check{g}_{\infty,L} \otimes \check{v}_{LR}  \right\rangle_{\hat{p}_F} \nonumber \\
& & + \left\langle \check{v}_{RL}  \otimes \check{g}_{\infty,L} \otimes \check{v}_{LR} \otimes  \check{g}_{\infty,R} \otimes \check{t}_{RR} \right\rangle_{\hat{p}_F} 
\end{eqnarray}
and
\begin{eqnarray}
\check{t}_{LR}  & = & \check{v}_{LR} + \left\langle \check{v}_{LR}  \otimes \check{g}_{\infty,R} \otimes \check{v}_{RL} \otimes  \check{g}_{\infty,L} \otimes \check{t}_{LR} \right\rangle_{\hat{p}_F} \\
\check{t}_{RL}  & = &  \check{v}_{RL}  + \left\langle \check{v}_{RL}  \otimes \check{g}_{\infty,L} \otimes \check{v}_{LR} \otimes  \check{g}_{\infty,R} \otimes \check{t}_{RL} \right\rangle_{\hat{p}_F} 
\end{eqnarray}
where the Fermi surface average $\left\langle \cdots \right\rangle_{\hat{p}_F}$ has to be performed with respect to each different Fermi surface sheet under consideration. If we now take a look at the quasiclassical transport equation for each band $\alpha$ where the strong perturbation at the interface enters as a source term on the right hand side
\begin{eqnarray}
i \vec{v}_F^{\alpha} \vec{\nabla} \check{g}_i^{\alpha} (\hat{p}_F) + \left[ \check{\epsilon}_i  (\hat{p}_F) -\check{\Delta}_i (\hat{p}_F) , \check{g}_i (\hat{p}_F) \right]_\otimes^{\alpha} \nonumber \\
= 2 \pi v_{F,n}^{\alpha} \left[ \check{t}_{ii} (\hat{p}_F,  \hat{p}_F ), \check{g}_{\infty, i} (\hat{p}_F)  \right]_\otimes^{\alpha} \delta(\vec{r} -\vec{r}_c) 
\label{eq:transporteq}
\end{eqnarray}
complemented by the normalization condition
\begin{equation}
\check{g} \otimes \check{g} = - \pi^2 \check{1}
\label{eq:normaliz}
\end{equation}
we can  calculate the rapid change of the quasiclassical propagator at the interface by integrating over the source term in Eq.~(\ref{eq:transporteq}) neglecting all the slowly varying fields on the left hand side. We get for the difference of the quasiclassical Green's function on the incoming and the outgoing trajectory -- evaluated directly at the boundary 
\begin{eqnarray}
& & \left[ \check{g}_{i,+} (\hat{p}_F) - \check{g}_{i,-} (\hat{p}_F) \right]^{\alpha} \nonumber \\
& & = - 2 \pi i
\left[ \check{t}_{ii} (\hat{p}_F,  \hat{p}_F ), \check{g}_{\infty, i} (\hat{p}_F)  \right]_\otimes^{\alpha} 
\label{eq:goutmgin}
\end{eqnarray}
Now it is possible to construct a solution for $\check{g}_{i,-}$ and $\check{g}_{i,+}$ that fulfills the boundary condition given by Eq.~(\ref{eq:goutmgin}) and that also satisfies the normalization condition of the quasiclassical propagator given by Eq.~(\ref{eq:normaliz}):
\begin{eqnarray}
\check{g}_{i,-}  & = & \check{g}_{\infty,i} + \left(\check{g}_{\infty,i} + i \pi \check{1} \right) \otimes \check{t}_{ii} \otimes \left(\check{g}_{\infty,i} - i \pi \check{1} \right)  \label{eq:gm} \\
\check{g}_{i,+}  & = & \check{g}_{\infty,i} + \left(\check{g}_{\infty,i} - i \pi \check{1} \right) \otimes \check{t}_{ii} \otimes \left(\check{g}_{\infty,i} + i \pi \check{1} \right) \label{eq:gp}
\end{eqnarray}
where we have omitted the angular dependencies. Taking $\check{g}_{i,\pm}$ as boundary values we can now construct the full quasiclassical propagator on incoming and outgoing trajectories in the vicinity of an interface with arbitrary transparency. In the following we will use it to calculate the quasiparticle density of states at the boundary, the Josephson currents across the boundary and the currents at an applied voltage for a multiband $s$-wave superconductor. Since there are no Andreev bound states at the interface without a sign change of the order parameter for reflected trajectories we do not expect dramatic changes of the pairing potential amplitude near the boundary. In this case we can go on without a self-consistent calculation of the local pairing potential.

\section{\label{sec:current} Current with and without an applied voltage}

The current density can be calculated integrating the Keldysh component of the quasiclassical  propagator over the quasiparticle energy variable $\epsilon$. We can write the current component in the band $\alpha$ as
\begin{equation}
\vec{j}^{\alpha} (\vec{r},t) = e N_F^{(\alpha)} \int \frac{d \epsilon}{8 \pi i} \mathrm{Tr} \left\langle \vec{v}_F^{\alpha}  \hat{\tau}_3 \hat{g}^{K,\alpha} (\vec{r}, \epsilon, t)  \right\rangle_{\alpha}  
\end{equation}
In the following we want to calculate only the current component across the surface  on one side of the contact, e.g. the left side, so we can write
\begin{eqnarray}
j_{n,L}^{\alpha}  & = & e N_F^{(\alpha)} \int \frac{d \epsilon}{8 \pi i} \left\langle v_{F,n}^{\alpha} \mathrm{Tr} \left[ \hat{\tau}_3 \left( \hat{g}_{L,-}^{K} - \hat{g}_{L,+}^{K}  \right) \right]^{\alpha} \right\rangle_{\alpha, \phi_+}  \nonumber \\
& = & e N_F^{(\alpha)} \int \frac{d \epsilon}{4} \left\langle v_{F,n}^{\alpha} j_\epsilon^{K,\alpha}   \right\rangle_{\alpha, +} \label{eq:jn}
\end{eqnarray}
where $\langle\cdots \rangle_{\alpha, +}$ means an average on the Fermi surface corresponding to the band $\alpha$ with the restriction that only angles corresponding to outgoing trajectories with $v_{F,n}^{\alpha} > 0$ are taken into account. As we have seen before we can calculate the difference of the incoming and the outgoing propagator in the vicinity of the surface as 
\begin{equation}
j_\epsilon^{K,\alpha} = \mathrm{Tr} \left\{\hat{\tau}_3 \left[ \check{t}_{LL} , \check{g}_{\infty, L}  \right]_\otimes^{K,\alpha} \right\}
\label{eq:spectral_current}
\end{equation}
This expression can be further simplified in the equilibrium case where the Keldysh propagator can be written as difference of the retarded and advanced Green's function multipled by a temperature dependent distribution function. In the non-equilibrium case where we apply a current over the contact we have to calculate the time evolution of the transfer-matrix and get a time-dependent solution for the current.  

\subsection{Josephson currents}
To calculate the currents we have to perform the Fermi surface average as defined in the previous section. Since within our model none of the Green's functions posses an angular dependence we only have to evaluate the following expression for the $\sigma$ and for the $\pi$ band
\begin{equation}
\left\langle v_{F,n}^{\sigma} \right\rangle_{\sigma, +}, \; \left\langle v_{F,n}^{\pi} \right\rangle_{\pi, +}
\end{equation}
To proceed further we must take into account that the different bands posses different Fermi surface geometries. As we have shown in a previous work the three dimensional tubular network of the $\pi$ bands can be approximatly described by a half torus with a fraction of the two toroidal radii of $\nu=4$ while the nearly two dimensional isotropic Fermi surfaces of the $\sigma$ bands can be described by distorted cylinders~\cite{dahm1,graser}. Taking the interface to be orientated perpendicular to the $ab$-plane direction we can write the Fermi surface averages as  
\begin{equation}
\left\langle v_F^{\sigma} \cos \phi \right\rangle_{\sigma, +} =  v_F^{\sigma} \frac{1}{2 \pi} \int_{-\pi/2}^{\pi/2} \cos \phi \; d \phi = \frac{1}{\pi} v_F^{\sigma}
\end{equation}
and 
\begin{eqnarray}
& & \left\langle v_F^{\pi} \cos \phi \cos \theta \right\rangle_{\pi, +} \nonumber \\
& = &  v_F^{\pi} \frac{1}{2 \pi} \int_{\pi/2}^{3 \pi/2} \int_{\pi/2}^{3 \pi/2} \cos \phi \frac{\nu + \cos \theta }{\pi \nu - 2} \cos \theta \; d \phi d \theta \nonumber \\
& = & \frac{1}{\pi} \frac{4 \nu - \pi}{2 \pi \nu - 4} v_F^{\pi} 
\end{eqnarray}
If we determine the values of $v_F^{\sigma} = 4.4 \cdot 10^5 \frac{\mathrm m}{\mathrm s}$ and $v_F^{\pi} = 8.2 \cdot 10^5 \frac{\mathrm m}{\mathrm s}$ from band structure calculations~\cite{brinkman,dahm1} and insert the value of $\nu = 4$ we finally get the weighting factors for the two different bands as 
\begin{equation}
\left\langle v_F^{\sigma} \cos \phi \right\rangle_{\sigma, +} =  1.4 \cdot 10^5 \frac{\mathrm m}{\mathrm s}, 
\end{equation}
and
\begin{equation}
\left\langle v_F^{\pi} \cos \phi \cos \theta \right\rangle_{\pi, +} =  1.588 \cdot 10^5 \frac{\mathrm m}{\mathrm s}
\end{equation}
Now we can calculate the Josephson current as a function of the phase difference between the gaps on the left and on the right hand side of the junction from Eq.~(\ref{eq:jn}) taking the incoming and outgoing Green's functions from Eq.~(\ref{eq:gm}) and Eq.~(\ref{eq:gp}) and evaluating the quasiclassical $t$-matrix as defined in Eq.~(\ref{eq:tLL}). Since the Josephson current is an equilibrium property the $\otimes$-products reduce to simple matrix multiplications. To compare the quasiclassical results of the temperature dependence of the critical currents with a simplified model we can write down a two band analogon to the Ambegaokar-Baratoff formula. Here we have to consider the weighted sum of two single band Ambegaokar-Baratoff formulas using the temperature dependence of the pairing potentials calculated selfconsistently from the two band gap equation:
\begin{equation}
j_c(T) = \sum_{\alpha} e N_F^{(\alpha)} \left\langle v_{F,n}^{\alpha} \right\rangle_{\alpha, +} \mathcal{D}^\alpha \pi  \Delta^{(\alpha)} \tanh \left( \frac{\Delta^{(\alpha)}}{2 T} \right) \label{eq:ambegaokar}
\end{equation}
Here $\mathcal{D}^\alpha$ is an effective band dependent transmission coefficient. For low transparencies it can be calculated from the hopping amplitudes as $\mathcal{D}^\alpha = 4 \pi^2 \sum_{\alpha'} \left| v^{\alpha,\alpha'} \right|^2$.

\subsection{$S$-$I$-$S$- and $N$-$I$-$S$-contacts}
If we apply a constant voltage across an $S$-$I$-$S$ Josephson junction we will find that the phase difference increases linear according to the Josephson relation $\phi(t)=\phi_0 + \omega_0 t$ where the Josephson frequency is proportional to the applied voltage $\omega_0=2eV/\hbar$. This results in an oscillating current across the junction and is known as the ac-Josephson effect. In this case we have to solve the non-equilibrium time-dependent $t$-matrix equation. First of all we will simplify the Keldysh component of $\left[ \check{t}_{LL} , \check{g}_{\infty, L}  \right]_\otimes$ and we will show that the problem can be reduced to calculate only the retarded and advanced parts of the transmission components of the quasiclassical $t$-matrix $\hat{t}^{R/A}_{LR}$ and $\hat{t}^{R/A}_{RL}$. Then we will determine the time dependence of the transfer-matrix components mentioned above by expanding them into a Fourier series.  
As shown in the appendix the commutator can be transformed in such a way that we can write the spectral current density as 
\begin{eqnarray}
j_\epsilon^K & = & \mathrm{Tr} \left\{ \hat{\tau}_3 \left( \hat{v}_{LR} \otimes \hat{g}_{\infty,R}^R \otimes \hat{t}_{RL}^R \otimes \hat{g}_{\infty,L}^K \otimes \hat{t}_{LR}^A \otimes \hat{v}_{LR}^{-1} \right. \right. \nonumber \\
& & - \hat{v}_{RL}^{-1} \otimes \hat{t}_{RL}^R \otimes \hat{g}_{\infty,L}^K \otimes \hat{t}_{LR}^A \otimes \hat{g}_{\infty,R}^A  \otimes \hat{v}_{RL} \nonumber \\
& & +  \hat{t}_{LR}^R \otimes \hat{g}_{\infty,R}^K \otimes \hat{t}_{RL}^A \otimes \hat{g}_{\infty,L}^A  \nonumber \\
& & \left. \left. - \hat{g}_{\infty,L}^R \otimes \hat{t}_{LR}^R \otimes \hat{g}_{\infty,R}^K \otimes \hat{t}_{RL}^A  \right) \right\}
\label{eq:neqcurrent}
\end{eqnarray}
Now we have to find the time evolution of the transfer-matrix. Since there is no further complication from the multiband character of our problem we can directly follow the steps described by Cuevas and Fogelstr\"om~\cite{cuevas2}. Starting with the transfer-matrix equation \begin{eqnarray}
\hat{t}_{LR} ^{R/A} (t,t') & = & \hat{v}_{LR}  +  \int dt_1 \int dt_2 \hat{v}_{LR}  \hat{g}_{\infty,R}^{R/A} (t,t_1) \nonumber \\
& & \times   \hat{v}_{RL}   \hat{g}_{\infty,L}^{R/A} (t_1,t_2) \hat{t}_{LR}^{R/A} (t_2,t') 
\end{eqnarray}
and Fourier transforming it to energy space  
\begin{equation}
\hat{t}_{LR}^{R/A} (t,t') = \frac{1}{2\pi} \int d\epsilon \int d\epsilon' e^{-i \epsilon t} e^{i \epsilon t'} \hat{t}_{LR}^{R/A} (\epsilon, \epsilon')
\end{equation}
leads to an algebraic equation to determine $\hat{t}_{LR} (\epsilon, \epsilon')$. Further one can show that  the $t$-matrix allows a Fourier expansion of the form 
\begin{equation}
\hat{t}_{LR}^{R/A} (\epsilon, \epsilon') = \sum_n \hat{t}_{LR}^{R/A} (\epsilon, \epsilon + neV) \delta(\epsilon - \epsilon' + neV)
\end{equation}
So the problem of calculating the current can be reduced to the problem of calculating the Fourier components $\hat{t}_{nm}$ for the retarded and advanced part of the transfer-matrix, that are equally defined as 
\begin{equation}
\hat{t}_{nm} = \hat{t}_{LR}^{R/A} (\epsilon + neV, \epsilon + meV)
\end{equation}
In the following we will use  the same notation for the advanced and the retarded part of the $t$-matrix since they are described by the same equation. We will also use the notation $v=v_{LR}$ and $v^\dagger=v_{RL}$. But we have to keep in mind, that all $\hat{t}_{nm}$ are not only 2$\times$2-matrices in Nambu-Gor'kov space but are also $N \times N$-matrices in band space. Finally we can write down the algebraic $t$-matrix equation as~\cite{cuevas2}
\begin{eqnarray}
\hat{t}_{n,m} & = & \hat{\nu}_{n,m} \delta_{n,m\pm 1} + \hat{\mathcal{E}}_{n,n} \hat{t}_{n,m} \nonumber \\
& & + \hat{\mathcal{V}}_{n,n-2} \hat{t}_{n-2,m} + \hat{\mathcal{V}}_{n,n+2} \hat{t}_{n+2,m} \label{eq:algebraic}
\end{eqnarray}
where we have defined the matrix components as
\begin{equation}
\hat{\mathcal{E}}_{n,n} = \left( \begin{array}{cc}
v g_{R,n+1} v^\dagger g_{L,n}  & v g_{R,n+1} v^\dagger f_{L,n}  \\
v^\dagger  g_{R,n-1} v \tilde{f}_{L,n}  & v^\dagger g_{R,n-1} v g_{L,n} 
\end{array} \right)
\end{equation}
and also
\begin{equation}
\hat{\mathcal{V}}_{n,n+2} = - v f_{R,n+1} v \left( \begin{array}{cc}
\tilde{f}_{L,n+2}  &  g_{L,n+2}  \\
0  & 0 
\end{array} \right)
\end{equation}
and finally
\begin{equation}
\hat{\mathcal{V}}_{n,n-2} = - v^\dagger \tilde{f}_{R,n-1} v^\dagger \left( \begin{array}{cc}
0  & 0  \\
g_{L,n-2}  &  f_{L,n-2}  
\end{array} \right)
\end{equation}
Here and in the following the notation $g_{i,n}$ stands for the diagonal and the functions $f_{i,n}$ and $\tilde{f}_{i,n}$ stand for the off-diagonal parts of the equilibrium Green's function, evaluated at $\epsilon + neV$ as 
\begin{equation}
\hat{g}_{i,n} = \hat{g}_{\infty,i} (\epsilon + neV)
\end{equation}
The matrices $\nu_{nm}$ are only evaluated for $n=m\pm1$ and we can write them as 
\begin{equation}
\hat{\nu}_{m-1,m} =  \left( \begin{array}{cc}
v  & 0  \\
0  &  0  
\end{array} \right), \hat{\nu}_{m+1,m} =  \left( \begin{array}{cc}
0  & 0  \\
0  & v^\dagger  
\end{array} \right)
\end{equation}
Now the algebraic set of equations defined by Eq.~(\ref{eq:algebraic}) can be solved by standard recursive techniques that we will briefly demonstrate. A significant simplification can be achieved if we use the following identity 
\begin{equation}
\hat{t}_{n,m}(\epsilon) = \hat{t}_{n-m,0}(\epsilon + meV)
\end{equation}  
that follows directly from the definition of the $\hat{t}_{n,m}(\epsilon)$. Now we can proceed further by defininig some ladder operators $\hat{z}_n^{\pm}$ as
\begin{equation}
\hat{t}_{n,0} = \hat{z}_n^{\pm} \hat{t}_{n \mp 2,0}
\end{equation}
It can be easily verified that $\hat{t}_{0,0}=\hat{0}$ is a solution of Eq.~(\ref{eq:algebraic}) for $n=0$ and therefore all $\hat{t}_{n,0}$ with even $n$ vanish identically. So we only have to calculate the odd ladder operators. Inserting the above relations in Eq.~(\ref{eq:algebraic}) leads to recursion formulas for the $\hat{z}_n^{\pm}$ for $n>1$ or $n<-1$ respectively:
\begin{equation}
\hat{z}_n^{\pm} = - \left( \hat{\mathcal{E}}_{n,n} - \hat{1} +\hat{\mathcal{V}}_{n,n \pm 2} \hat{z}_{n\pm 2}^\pm \right)^{-1} \hat{\mathcal{V}}_{n,n \mp 2}
\end{equation}
Since the $\hat{t}_{n,0}$ are getting constant for high values of $n$ we can start  with $\hat{z}_{n_{max}}^{+} = 0$ and $\hat{z}_{-n_{max}}^{-} = 0$ counting down to $\hat{z}_3^{+}$ or up to $\hat{z}_{-3}^{-}$. After solving the equations for $\hat{t}_{\pm 1,0}$ we are now able to determine all $\hat{t}_{n,0}$ by recursive application of the ladder operators. To find the $\hat{t}_{\pm 1,0}$ we have to solve two coupled $6 \times 6$ matrix equations in the combined band and Nambu-Gorkov space:
\begin{eqnarray}
\hat{t}_{1,0} & = & \hat{\nu}_{1,0} + \hat{\mathcal{E}}_{1,1} \hat{t}_{1,0} + \hat{\mathcal{V}}_{1,-1} \hat{t}_{-1,0} + \hat{\mathcal{V}}_{1,3} \hat{z}_3^+ \hat{t}_{1,0}  \\
\hat{t}_{-1,0} & = & \hat{\nu}_{-1,0} + \hat{\mathcal{E}}_{-1,-1} \hat{t}_{-1,0} + \hat{\mathcal{V}}_{-1,-3} \hat{z}_{-3}^- \hat{t}_{-1,0} + \hat{\mathcal{V}}_{-1,1} \hat{t}_{1,0}  \nonumber
\end{eqnarray} 
Of course this procedure has to be applied to the retarded as well as to the advanced components of the transfer matrices. The missing transfer matrices can be found using the general relation $\hat{t}_{RL,nm}^{A/R} (\epsilon) = \hat{\tau}_3 \hat{t}_{LR,mn}^{R/A \dagger} (\epsilon) \hat{\tau}_3$ as for example described by Cuevas and Fogelstr\"om~\cite{cuevas2}. With this procedure the spectral weight of the non-equilibrium current across an $S$-$I$-$S$ contact for a given voltage $V$ and a given quasiparticle energy $\epsilon$ can be calculated. From this we get the time dependent non-equilibrium current in a voltage biased contact as
\begin{equation}
j_{n,L}^{\alpha}(V,t) = e N_F^{(\alpha)} \left\langle v_F^\alpha \right\rangle_{\alpha,+} \sum_{m=-\infty}^{\infty} j_m e^{2mieVt/\hbar}
\end{equation} 
with the Fourier components
\begin{eqnarray}
j_m & = &  \int \frac{d \epsilon}{4} \sum_n \mathrm{Tr} \left\{ \hat{\tau}_3 \left( \hat{v}_{LR} \hat{g}_{R,0}^R \hat{t}_{RL,0n}^R  \hat{g}_{L,n}^K \hat{t}_{LR,nm}^A \hat{v}_{LR}^{-1} \right. \right. \nonumber \\
& & - \hat{v}_{RL}^{-1} \hat{t}_{RL,0n}^R \hat{g}_{L,n}^K \hat{t}_{LR,nm}^A \hat{g}_{R,m}^A  \hat{v}_{RL}  \\
& & + \left. \left. \hat{t}_{LR,0n}^R \hat{g}_{R,n}^K \hat{t}_{RL,nm}^A \hat{g}_{L,m}^A  - \hat{g}_{L,0}^R \hat{t}_{LR,0n}^R \hat{g}_{R,n}^K \hat{t}_{RL,nm}^A \right) \right\} \nonumber
\end{eqnarray}
For the special case  of an $N$-$I$-$S$ contact we can choose without loss of generality the right hand side as the normal conducting side with $\Delta_R^{(\alpha)} = 0$ and therefore also $\hat{\mathcal{V}}_{n,n \pm 2} = \hat{0}$. In this case the ladder operators vanish and we only have to calculate the transfer matrix components $\hat{t}_{\pm 1,0}$. This can be understood since the ladder operators create repeated Andreev reflections between the superconducting electrodes and in the case of a superconductor-normal conductor interface we only have single Andreev processes.

\section{\label{sec:hoppingampl} Determination of the effective hopping amplitudes for MgB$_2$}
\begin{figure}[t]
  \begin{center}
     \includegraphics[width=0.9\columnwidth]{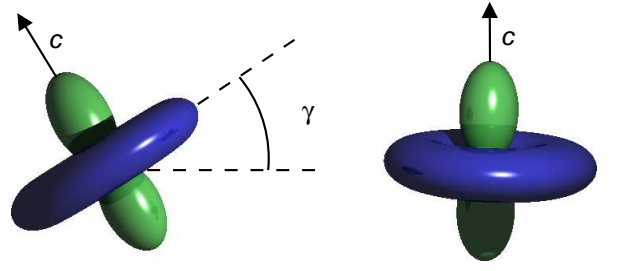}
  \end{center}
   \caption{(color online). The effective hopping amplitudes between the different bands of MgB$_2$ can be modelled by calculating the overlap integrals of the $p_x \pm i p_y$-orbitals and the $p_z$-orbitals of the boron atoms, which are responsible for the formation of the superconducting $\sigma$ and $\pi$ bands of this material.  The angle $\gamma$ measures the relative orientation of the two crystals on the different sides of the interface. \label{fig:orbitals}}
\end{figure} 
One of the crucial features of the two gap superconductor MgB$_2$ is the orthogonality of the wavefunctions belonging to the $\pi$ and $\sigma$ bands, respectively. Here, we want to consider the case that the two MgB$_2$ electrodes possess a certain misorientation angle of the crystal $c$-axis directions on both sides of the junction. The misorientation angle allows to change the interband transmission between the two sides. In the following we will introduce a simple model to determine the effective hopping amplitudes at the interface between these two misaligned regions. The $\sigma$ bands are formed by the in-plane $\sigma$ bonds of the $p_x \pm i p_y$ orbitals of the boron atoms, while the $\pi$ band is formed by the weaker $\pi$ bonds of the boron $p_z$ orbitals connecting the different boron layers. To model an interface between two electrodes with different orientation we have to calculate the overlap integrals between the different orbitals rotated against each other (see Fig. \ref{fig:orbitals}). Since all orbitals show rotational symmetry with respect to the $z$-axis it is sufficient to study only the rotation around one of the other axes, e.g. the $x$-axis. We can write down the effective one-particle wave function for the $p_x$, the $p_y$ and the $p_z$ orbitals, neglecting the radial part of the wave function, as follows
\begin{eqnarray}
\left| p_x \right\rangle & = & -\frac{1}{\sqrt{2}} \left(Y_{1,1} (\theta,\phi) - Y_{1,-1} (\theta,\phi) \right) \\
\left| p_y \right\rangle & = & \frac{i}{\sqrt{2}} \left(Y_{1,1} (\theta,\phi) + Y_{1,-1} (\theta,\phi) \right) \\
\left| p_z \right\rangle & = & Y_{1,0} (\theta,\phi) 
\end{eqnarray}
where $Y_{l,m} (\theta,\phi)$ are spherical harmonics. From this we can construct the $p_{xy}$ orbitals as 
\begin{equation}
\left| p_{xy}^\pm \right\rangle =  \frac{1}{\sqrt{2}} \left(\left| p_x \right\rangle \pm i \left| p_y \right\rangle \right) = \mp Y_{1,\pm1} (\theta,\phi) 
\end{equation}
which correspond to the two $\sigma^+$ and $\sigma^-$ sub-bands, as we will call them in the following. If we introduce rotated spherical coordinates we can define the following set of transformations between the unrotated and the rotated frame
\begin{eqnarray}
\sin \theta' \cos \phi' & = & \sin \theta \cos \phi \\
\sin \theta' \sin \phi' & = & \cos \gamma \sin \theta \sin \phi + \sin \gamma \cos \theta \\
\cos \theta' & = & - \sin \gamma \sin \theta \sin \phi + \cos \gamma \cos \theta
\end{eqnarray}
where $\gamma$ is the misorientation angle (see Fig. \ref{fig:orbitals}). Now we can calculate the overlap integrals for the different orbitals as
\begin{eqnarray}
\left\langle p_z | p_z (\gamma) \right\rangle & = & \left\langle Y_{1,0}^* (\theta,\phi) Y_{1,0} (\theta',\phi') \right\rangle_\Omega \nonumber \\
& = & \cos \gamma \\
\left\langle p_{xy}^\pm | p_z (\gamma) \right\rangle & = & \mp \left\langle Y_{1,\pm 1}^* (\theta,\phi) Y_{1,0} (\theta',\phi') \right\rangle_\Omega \nonumber \\
& = & \pm \frac{i}{\sqrt{2}} \sin \gamma \\
\left\langle p_{xy}^\pm | p_{xy}^\pm (\gamma) \right\rangle & = & \left\langle Y_{1,\pm 1}^* (\theta,\phi) Y_{1,\pm 1} (\theta',\phi') \right\rangle_\Omega \nonumber \\
& = & \frac{1}{2} \left( 1+ \cos \gamma \right) \\
\left\langle p_{xy}^\mp | p_{xy}^\pm (\gamma) \right\rangle & = & - \left\langle Y_{1,\mp 1}^* (\theta,\phi) Y_{1,\pm 1} (\theta',\phi') \right\rangle_\Omega \nonumber \\
& = & \frac{1}{2} \left( 1- \cos \gamma \right)
\end{eqnarray}
where we have used the abbreviation $\left\langle \dots \right\rangle_\Omega = \int_0^\pi d\theta \int_0^{2\pi} d\phi \sin \theta \dots$. The missing integrals can be found by simple symmetry considerations. If we now calculate the total transmission probability for one particle for example from the $\sigma^+$-band into one of the other bands we get
\begin{eqnarray}
& & \left| \left\langle p_{xy}^+ | p_{xy}^+ (\gamma) \right\rangle \right|^2 + \left| \left\langle p_{xy}^+ | p_{xy}^- (\gamma) \right\rangle \right|^2 + \left| \left\langle p_{xy}^+ | p_z (\gamma) \right\rangle \right|^2 \nonumber \\
& & = \frac{1}{4} (1 + \cos \gamma)^2 +\frac{1}{4} (1 - \cos \gamma)^2 + \frac{1}{2} \sin^2 \gamma = 1
\end{eqnarray}
as to be expected. We find the same total transmission propability for the other bands as well. If we multiply the overlap integrals with a global transmission factor $t$ describing the overall reflection and transmission properties of the point contact we can write the angular dependent effective hopping amplitudes as
\begin{eqnarray}
& & \tau_{\sigma^+ \sigma^+}  = \tau_{\sigma^- \sigma^-}  = \frac{t}{2} ( 1 + \cos \gamma) \\
& & \tau_{\sigma^+ \sigma^-} = \tau_{\sigma^- \sigma^+} = \frac{t}{2} ( 1 - \cos \gamma) \\
& & \tau_{\sigma^+ \pi} = \tau_{\sigma^- \pi} = \tau_{\pi \sigma^+} = \tau_{\pi \sigma^-} = \frac{t}{\sqrt{2}} \sin \gamma \\
& & \tau_{\pi \pi} =  t \cos \gamma 
\end{eqnarray}
We can write them in matrix form as 
\begin{equation}
\check{v}_{LR} = \left( \begin{array}{ccc}
\tau_{\sigma^+ \sigma^+} & \tau_{\sigma^+ \sigma^-} & \tau_{\sigma^+ \pi} \\
\tau_{\sigma^- \sigma^+} & \tau_{\sigma^- \sigma^-} & \tau_{\sigma^- \pi} \\
\tau_{\pi \sigma^+} & \tau_{\pi \sigma^-} & \tau_{\pi \pi}
\end{array} \right) \check{1}
\end{equation}
The effective hopping amplitudes enter the transfer-matrix equation as described in section \ref{sec:tmapproach}. 
\begin{figure}[t]
  \begin{center}
     \includegraphics[width=0.8 \columnwidth]{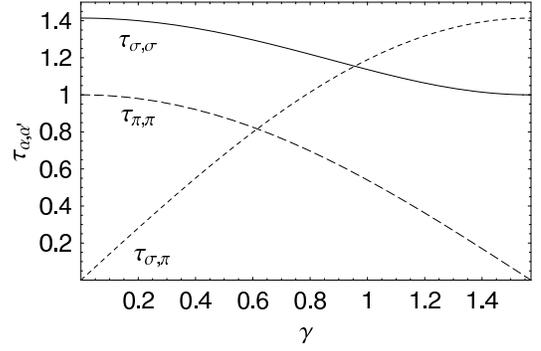}
  \end{center}
   \caption{The effective hopping amplitudes between the different bands as functions of the misorientation angle $\alpha$. Here we have added the contributions from the two different $\sigma$ bands defining $\tau_{\sigma \sigma} =\left[ 2 \left( \tau_{\sigma_+  \sigma_+}^2 +\tau_{\sigma_+ \sigma_-}^2 \right) \right]^{1/2} $, $\tau_{\sigma \pi} = 2 \tau_{\sigma_+  \pi} $ and $\tau_{\pi \pi}$ as before. \label{fig:transmission}}
\end{figure} 
In Fig.~\ref{fig:transmission} we show the angular dependence of the effective hopping amplitudes. Note, that interband transmission vanishes for $\gamma=0$, when there is no misalignment. On the other hand, for $\gamma =\pi/2$ interband transmission becomes maximal, however, intraband transmission from the $\pi$ band into the $\pi$ band $\tau_{\pi \pi}$ vanishes. This peculiar behavior is a direct consequence of the orthogonality of the $\sigma$ and $\pi$ bands.

\section{\label{sec:results} Results}
In this section we will discuss the dependence of the local density of states, the critical current and the differential conductance on the misorientation angle $\gamma$ between the two electrodes of a superconducting MgB$_2$ point contact. In the following we assume that the gap value in the vicinity of the point contact is not changed so it can be taken as the bulk value. All energies will be normalized to the larger gap value $\Delta^{(\sigma)}$ while we will take the ratio of the two gaps to be $\Delta^{(\pi)} = \frac{1}{3} \Delta^{(\sigma)}$.~\cite{liu}  
\begin{figure}[t]
  \begin{center}
     \includegraphics[width=0.8\columnwidth]{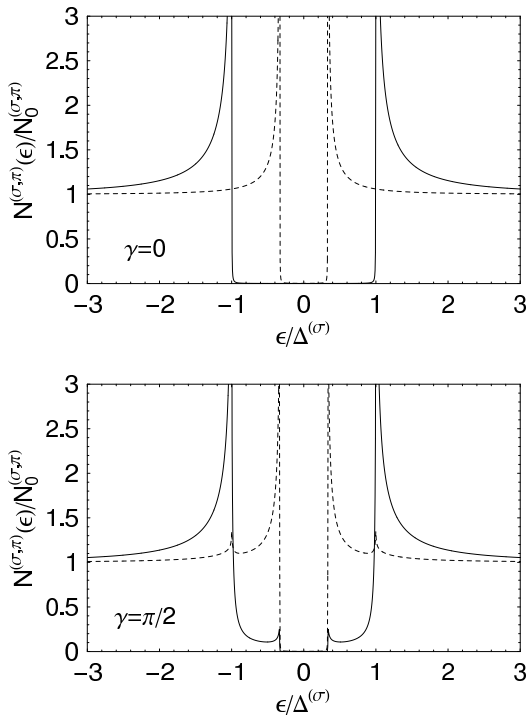}
  \end{center}
   \caption{The quasiparticle density of states at a point contact calculated within a two band model. The full line shows the DOS of the quasiparticles in the $\sigma$ bands while the dashed line shows the DOS of the quasiparticles in the $\pi$ band. Both curves are normalized to the normal conducting DOS in the corresponding bands. \label{fig:densityofstates}}
\end{figure} 
First we will calculate the local quasiparticle density of states (LDOS) in the vicinity of the point contact. In equilibrium without an applied voltage we can calculate the LDOS in band $\alpha$ as 
\begin{equation}
N_i^{(\alpha)} (\epsilon) = -\frac{N_F^{(\alpha)}}{2 \pi i} \left( \left\langle g_{i,+}^{R,\alpha} (\epsilon) \right\rangle_{\alpha,+} + \left\langle g_{i,-}^{R,\alpha} (\epsilon) \right\rangle_{\alpha,-} \right)
\end{equation}
using the expressions of Eqs.~(\ref{eq:gm}) and (\ref{eq:gp}) to calculate the quasiclassical Green's function on the incoming and the outgoing trajectory. In Fig.~\ref{fig:densityofstates} the LDOS is shown for two different contacts. In the first case we assume that both electrodes are equally orientated ($\gamma = 0$) and we have no interband tunneling. Then the different bands remain decoupled and we find a standard $s$-wave spectrum in the $\sigma$ bands as well as in the $\pi$ band.  In the second case we assume a maximum misorientation ($\gamma=\frac{\pi}{2}$) and we find peaks at the position of the $\pi$ gap in the $\sigma$ band spectrum and vice versa due to interband hopping processes. In both cases the transparency parameter $t$ was chosen to be $t=0.1$. 
\begin{figure}[t]
  \begin{center}
     \includegraphics[width=0.8\columnwidth]{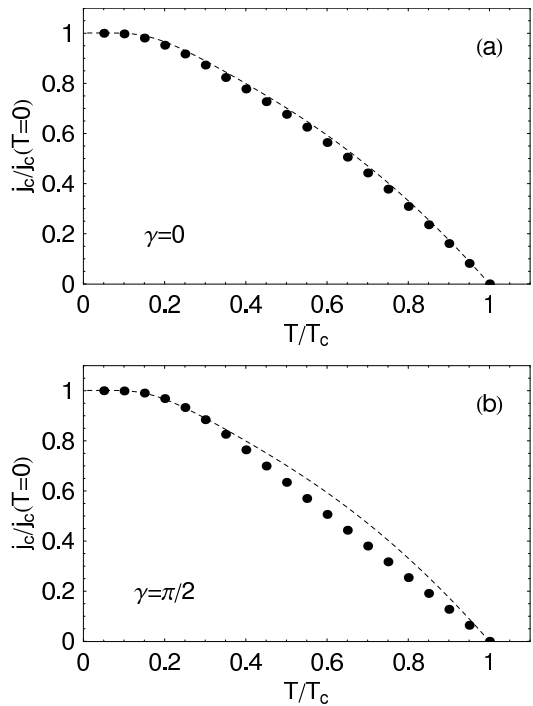}
  \end{center}
   \caption{Critical current as a function of temperature calculated within the transfer-matrix approach for two different orientations of the   
    electrodes (filled points). The dashed line gives the corresponding two-band Ambegaokar-Baratoff approximation. \label{fig:criticalc}}
\end{figure} 
In Fig.~\ref{fig:criticalc} we present calculations of the temperature dependence of the critical current for the same parameters. For two equally orientated electrodes we find a very good agreement of the quasiclassical results for low transparencies with the analytical approximation given by Eq.~(\ref{eq:ambegaokar}). This is clear since for $\gamma=0$ we have a decoupled system and the critical current can be understood as a weighted sum of two single gap critical currents, that are well described by the Ambegaokar-Baratoff formula. For a maximum misorientation of $\gamma = \frac{\pi}{2}$ we find a deviation from the classical behaviour and a nearly linear decrease of the critical current over a wide range of temperatures. In both cases we have taken the normalized density of states at the Fermi energy to be $N_F^{(\sigma)} = \frac{0.7}{1.7}$ and $N_F^{(\pi)} = \frac{1}{1.7}$ corresponding to band structure calculations for MgB$_2$.~\cite{liu} The deviation from the Ambegaokar-Baratoff result seen in Fig.~\ref{fig:criticalc}b can be attributed to interband transmission processes.
\begin{figure}[t]
  \begin{center}
     \includegraphics[width=0.8\columnwidth]{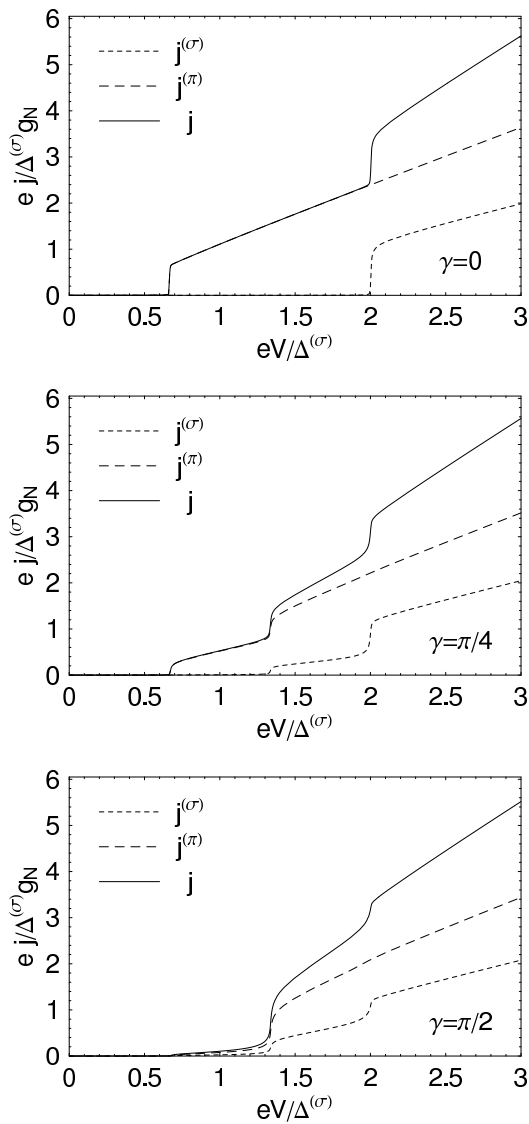}
  \end{center}
   \caption{Current as function of applied voltage for an $S$-$I$-$S$ junction calculated in a multiband model   
    with parameters adapted to MgB$_2$ (e.g. $\Delta^{(\sigma)} = 3 \Delta^{(\pi)}$) and low transparency. The angle
   $\gamma$ gives the relative orientation of the two crystals on both sides of the interface and determines the effective strength of interband
   and intraband tunneling. \label{fig:current_lowt}}
\end{figure}  
\begin{figure}[t]
  \begin{center}
     \includegraphics[width=0.8\columnwidth]{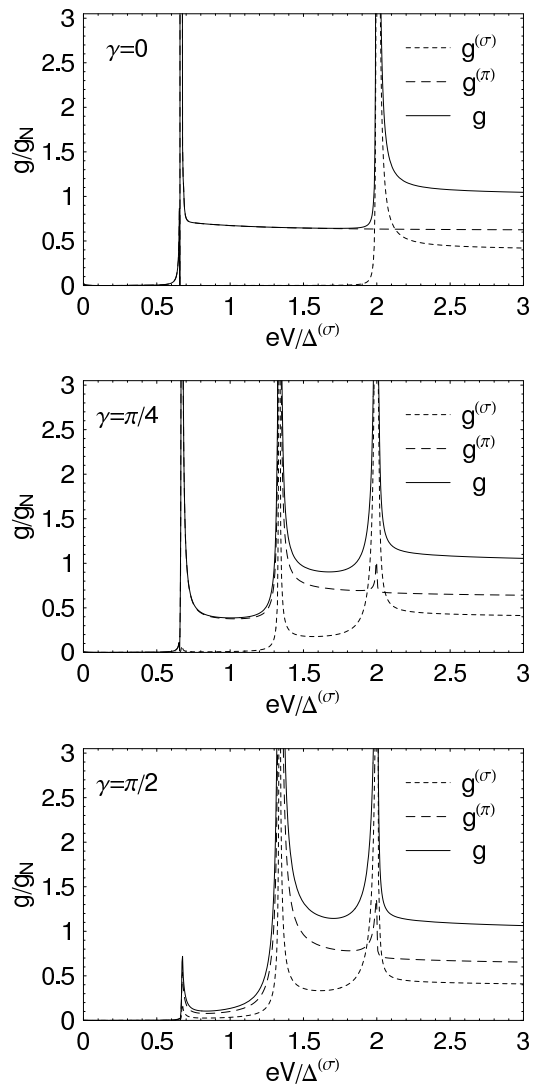}
  \end{center}
  \caption{Differential conductance as function of applied voltage for an $S$-$I$-$S$ junction calculated in a multiband model   
    with parameters adapted to MgB$_2$ (e.g. $\Delta^{(\sigma)} = 3 \Delta^{(\pi)}$) and low transparency. The angle
   $\gamma$ gives the relative orientation of the two crystals on both sides of the interface and determines the effective strength of     
    interband and intraband tunneling. \label{fig:conductance_lowt}}
\end{figure}  
In the next figures we consider a point contact with an applied voltage. The current $j$ and the differential conductance $g$ can be normalized using the conductance of the contact in the normal state $g_N$. Employing Eq.~(\ref{eq:jn}) and Eq.~(\ref{eq:neqcurrent}) the normal state conductance for quasiparticles of the band $\alpha$ can be calculated as
\begin{equation}
g_N^\alpha = 4 \pi^2 e N_F^{(\alpha)} \left\langle v_{F,n}^\alpha \right\rangle_{\alpha,+}  \left[ v \left( 1 + \pi^2 v^2 \right)^{-1} v \left( 1 + \pi^2 v^2 \right)^{-1} \right]^\alpha
\end{equation} 
and the total conductance is a sum over all band components $g_N = \sum_\alpha g_N^\alpha$. For high transparencies it shows a strong dependence on the misorientation angle $\gamma$ while it becomes independent of $\gamma$ in the limit of low transparency. In this limit also the calculation of the non-equilibrium properties can be considerably simplified. In the case of low transparency we can neglect multiple Andreev reflections and use a simple model that can be deduced from Eq.~(\ref{eq:neqcurrent}) expanding it for small values of $v$  and keeping only terms in second order.  In this case only the $t$-matrix components $\hat{t}_{\pm 1,0}$ have to be calculated and they reduce to $\hat{t}_{\pm 1,0} = \hat{\nu}_{\pm 1,0}$. Then all the different terms in Eq.~(\ref{eq:neqcurrent}) can be summed up to 
\begin{eqnarray}
j_{n,L} & = &  4 \pi^2 \sum_{\alpha,\alpha'} e N_F^{(\alpha)} \left\langle v_{F,n}^\alpha \right\rangle_{\alpha,+} \left| v^{\alpha,\alpha'} \right|^2 \\
& & \times \int \frac{d \epsilon}{2} \; \rho_L^\alpha (\epsilon) \rho_R^{\alpha'} (\epsilon - eV) \left( f_F (\epsilon) -f_F(\epsilon - eV) \right) \nonumber 
\end{eqnarray}
where we have used that in equilibrium $g_j^K (\epsilon) = \left[ g_j^R (\epsilon)  - g_j^A (\epsilon)  \right] f_F (\epsilon)$ with a fermionic distribution function $f_F (\epsilon) = \tanh \frac{\epsilon}{2T}$. Furthermore we have written the quasiparticle density of states as $\rho_j^\alpha (\epsilon) = -\frac{1}{2 \pi i} \left[ g_j^{R,\alpha} (\epsilon) - g_j^{A,\alpha} (\epsilon) \right]$. In Fig.~\ref{fig:current_lowt} we have calculated  the current voltage characteristics for three different misorientation angles $\gamma$ within this model. The dotted and dashed lines show the contributions of the different bands where we have added the formally equivalent contributions from the two $\sigma$ bands. For two equally orientated crystals on both sides of the interface ($\gamma = 0$) we have no interband hopping amplitudes and therefore we find two distinct increases in the current stemming from the $\sigma$ band contribution at $V = 2 \Delta^{(\sigma)}$ and from the $\pi$ band contribution at $V = 2 \Delta^{(\pi)}$. This can be seen as a simple addition of two decoupled single band contributions with different gap sizes. For $\gamma=\frac{\pi}{4}$ we find three distinct increases, the first one for $V=2 \Delta^{(\pi)}$, the second one for $V = \Delta^{(\pi)} + \Delta^{(\sigma)}$ and the last one for $V= 2 \Delta^{(\sigma)}$ since the intraband as well as the interband hopping amplitudes become finite. For $\gamma = \frac{\pi}{2}$ we find a maximal increase in the tunneling current for $V = \Delta^{(\pi)} + \Delta^{(\sigma)}$ and a smaller contribution at $V= 2 \Delta^{(\sigma)}$ as expected from the angle dependence of the hopping amplitudes. Intraband processes between the $\pi$ bands do not exist for this orientation since the $\frac{\pi}{2}$ rotated $p_z$ orbitals are orthogonal.  In Fig.~\ref{fig:conductance_lowt} we show the corresponding differential conductance as a function of voltage obtained by numerical differentiation of the current voltage characteristics in Fig.~\ref{fig:current_lowt}. Here the increase of the tunneling current due to tunneling between the high density of states at the gap edges results in distinct peaks in the differential conductance.
\begin{figure}[t]
  \begin{center}
     \includegraphics[width=0.8\columnwidth]{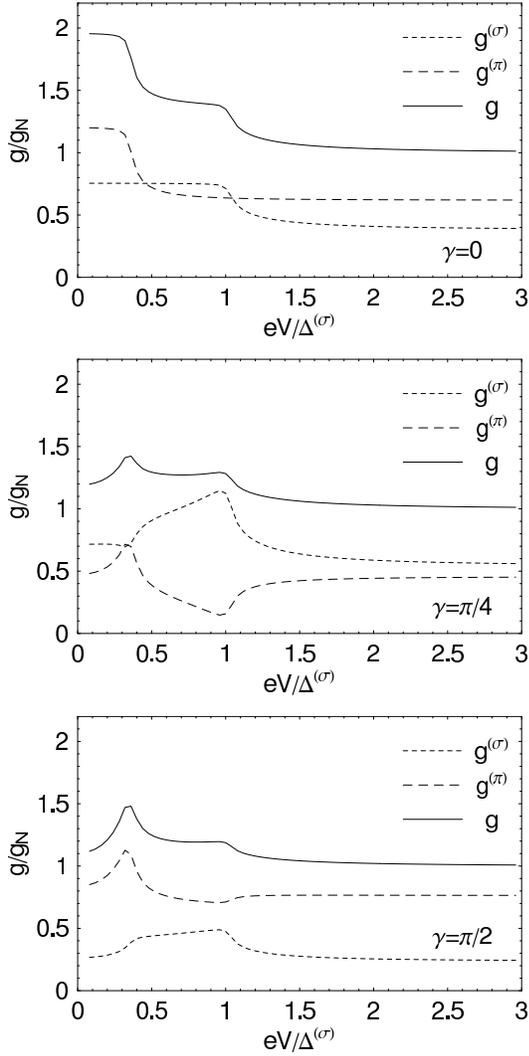}
  \end{center}
  \caption{Differential conductance as function of applied voltage for an $N$-$I$-$S$ junction calculated in a multiband model   
    with parameters adapted to MgB$_2$ (e.g. $\Delta^{(\sigma)} = 3 \Delta^{(\pi)}$) and maximal transparency $t=1/\pi$. The angle
   $\gamma$ gives the relative orientation of the two crystals on both sides of the interface and determines the effective strength of interband
   and intraband tunneling. \label{fig:current_NIS}}
\end{figure}
In comparison to Fig.~\ref{fig:current_lowt} and Fig.~\ref{fig:conductance_lowt} the situation for point contacts with high transparency is totally different. Here the occurrence of Andreev reflections dominates the current voltage characteristics. First we want to consider an $N$-$I$-$S$ junction with multiband electrodes on both sides of the interface. This situation could for example be experimentally realized by a reduced critical temperature in one of the electrodes due to defects.  Andreev reflection processes carry twice the charge as the corresponding one electron process. Thus we would expect in the case of maximum transparency an increase of the differential conductance for low voltage leading to a value that is two times larger than the high voltage asymptotics. This case is realized in Fig.~\ref{fig:current_NIS} for $\gamma=0$ and the results are in good agreement with the BTK-results presented by Brinkman et al.~\cite{brinkman}. If we assume a misorientation of the electrodes we find a decrease of the differential conductance for low energies that can be explained by a reduction of the current in the $\pi$ band as a result of an effectively smaller transparency of the barrier. 
\begin{figure}[t]
  \begin{center}
     \includegraphics[width=0.8\columnwidth]{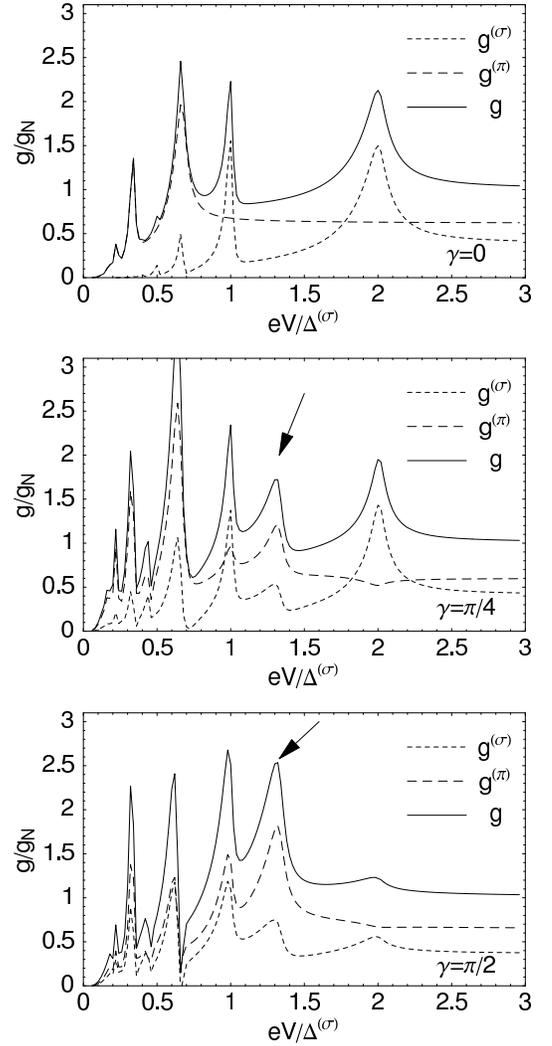}
  \end{center}
  \caption{Differential conductance as function of applied voltage for an $S$-$I$-$S$ junction calculated in a multiband model   
    with parameters adapted to MgB$_2$ (e.g. $\Delta^{(\sigma)} = 3 \Delta^{(\pi)}$) and high transparency $t=0.1$. The angle
   $\gamma$ gives the relative orientation of the two crystals on both sides of the interface and determines the effective strength of interband
   and intraband tunneling. \label{fig:conductance_hight}}
\end{figure}  

In the last figure (Fig.~\ref{fig:conductance_hight}) we show the differential conductance for an MgB$_2$ $S$-$I$-$S$ point contact calculated within the non-equilibrium quasiclassical theory. For $\gamma=0$ we find distinct conductance peaks at $V= 2 \Delta^{(\pi)}$
and $V= 2 \Delta^{(\sigma)}$ and also the well known subgap structures at $V =\frac{2 \Delta^{(\alpha)}}{n}$ with $n \in \mathbb{N}$ that appear due to multiple Andreev reflections. If we rotate the electrodes we find a more complicated situation. For $\gamma = \frac{\pi}{4}$ the peak at $V= 2 \Delta^{(\sigma)}$ shrinks visibly as expected due to a reduced intraband hopping probability while the peak at $V =  \frac{2}{3} \Delta^{(\sigma)} = 2 \Delta^{(\pi)}$ increases. This can be easily understood since not only hopping processes between the $\pi$ bands contribute to this peak but also multiple Andreev reflections corresponding to  $V = \frac{1}{3}  2 \Delta^{(\sigma)}$ and interband processes with $V = (\Delta^{(\sigma)} + \Delta^{(\pi)} ) / 2$. For $\gamma = \frac{\pi}{2}$ the peak at $V= 2 \Delta^{(\sigma)}$ has nearly vanished while we find at $V= 2 \Delta^{(\pi)}$ a dramatic reduction of the differential conductance. Instead we find an increase of interband processes corresponding to $V= \Delta^{(\sigma)} +  \Delta^{(\pi)} = \frac{4}{3} \Delta^{(\sigma)}$ (arrows). This peak at $\Delta^{(\sigma)}+\Delta^{(\pi)}$ is a characteristic signature of interband transmission. It is absent for $\gamma=0$, when the electrodes are aligned, and it becomes maximum for $\gamma=\frac{\pi}{2}$. An experimental observation of this effect would provide direct evidence for the orthogonality of the wavefunctions belonging to the $\pi$ and $\sigma$ bands.

\section{\label{sec:conclusion} Conclusions}
Due to its multiband character MgB$_2$ Josephson junction promise to show rich and interesting physical properties. We presented calculations of current-voltage characteristics of both $S$-$I$-$S$ and $N$-$I$-$S$ contacts. We developed a simple model to determine the effective intraband and interband hopping amplitudes at the interface from a microscopic  point of view. We used these model parameters within a multiband transfer-matrix description based on the work of Cuevas and Fogelstr\"om and Kopu~et~al. to study the dependence of the tunneling currents on the relative orientation of the two electrodes. Especially due to the effect of multiple intraband and interband Andreev reflections we were able to identify dramatic changes in the differential conductance for different electrode orientation. For $N$-$I$-$S$ junctions with equally orientated electrodes our result agrees with the BTK-results of Brinkman~et~al. For $S$-$I$-$S$ junctions we found subgap structures at different voltages corresponding to different combinations of the two gaps. The features sensitively depend on the misorientation angle of the junction. We hope that especially the angular dependence of the differential conductance can be experimentally verified for point contacts on MgB$_2$ single crystals which would provide direct evidence for the orthogonality of the two bands. 

\acknowledgments
We would like to thank N.~Schopohl, C.~Iniotakis and A.~Gumann for valuable discussions. This work was supported through the 'Projektf\"orderung f\"ur Nachwuchswissenschaftler' of the University of T\"ubingen.

\appendix

\section{Calculation of the current expression}
\label{sec:app1}
In this appendix we will show how the commutator formula for the spectral current density $j_\epsilon^{K}$ can be transformed into an expression that only contains the retarded and advanced components of the off-diagonal transfer matrix elements $\check{t}_{LR}$ and $\check{t}_{RL}$. 

\subsection{Relation between the Keldysh and the retarded and advanced components}
\label{sub:app11}
First we show that the Keldysh component of the $T$-matrix can be replaced by the corresponding advanced and retarded components. In the following we will show the relations for the Gor'kov Green's functions but of course they are also valid for the components of the quasiclassical  Green's propagator. Starting with Eq.~(\ref{eq:t-matrix}), writing it out in Keldysh-space we get for the retarded and the Keldysh components of the $T$-matrix the following relations
\begin{equation}
\tilde{\hat{\mathcal{T}}}^R = \tilde{\hat{v}} + \tilde{\hat{v}} \circ  \tilde{\hat{\mathcal{G}}}_\infty^R \circ \tilde{\hat{\mathcal{T}}}^R
\end{equation} 
and
\begin{equation}
\tilde{\hat{\mathcal{T}}}^K = \tilde{\hat{v}} \circ \left(   \tilde{\hat{\mathcal{G}}}_\infty^R \circ \tilde{\hat{\mathcal{T}}}^K + \tilde{\hat{\mathcal{G}}}_\infty^K \circ \tilde{\hat{\mathcal{T}}}^A \right)
\end{equation} 
Solving the first equation for $\tilde{\hat{\mathcal{T}}}^R$ and the second equation for $\tilde{\hat{\mathcal{T}}}^K$ we get the following expressions
\begin{equation}
\tilde{\hat{\mathcal{T}}}^R = \left( \tilde{\hat{1}} - \tilde{\hat{v}} \circ  \tilde{\hat{\mathcal{G}}}_\infty^R \right)^{-1} \circ \tilde{\hat{v}}
\end{equation} 
and also
\begin{equation}
\tilde{\hat{\mathcal{T}}}^K = \left( \tilde{\hat{1}} - \tilde{\hat{v}} \circ  \tilde{\hat{\mathcal{G}}}_\infty^R \right)^{-1} \circ \tilde{\hat{v}} \circ \tilde{\hat{\mathcal{G}}}_\infty^K \circ \tilde{\hat{\mathcal{T}}}^A 
\end{equation}
Substituting the first equation into the second equation we end up with the important relation
\begin{equation}
\tilde{\hat{\mathcal{T}}}^K = \tilde{\hat{\mathcal{T}}}^R \circ \tilde{\hat{\mathcal{G}}}_\infty^K \circ \tilde{\hat{\mathcal{T}}}^A 
\end{equation}
With this equation we are able to replace the Keldysh component of the transfer matrix by the Keldysh component of the decoupled Green's function, that can be easily calculated within the quasiclassical theory.  

\subsection{Relations between the reflection and the transmission components of the transfer-matrix}
\label{sub:app12}
In a second step we will replace the reflection component $\check{\mathcal{T}}_{LL}$ by the corresponding transmission components $\check{\mathcal{T}}_{LR}$ and $\check{\mathcal{T}}_{RL}$. Comparing the expressions for the transmission and reflection components we find
\begin{eqnarray}
\check{\mathcal{T}}_{LL}  & = &  \left(\check{1} - \check{v}_{LR}  \circ \check{\mathcal{G}}_{\infty,R} \circ \check{v}_{RL}  \circ  \check{\mathcal{G}}_{\infty,L} \right)^{-1} \nonumber \\
& & \circ \check{v}_{LR}  \circ \check{\mathcal{G}}_{\infty,R} \circ \check{v}_{RL}  \\
\check{\mathcal{T}}_{LR}  & = &  \left(\check{1} - \check{v}_{LR}  \circ \check{\mathcal{G}}_{\infty,R} \circ \check{v}_{RL}  \circ  \check{\mathcal{G}}_{\infty,L} \right)^{-1} \circ \check{v}_{LR}  \nonumber
\end{eqnarray}
and therefore 
\begin{equation}
\check{\mathcal{T}}_{LL} = \check{\mathcal{T}}_{LR} \circ \check{\mathcal{G}}_{\infty,R} \circ \check{v}_{RL}
\label{eq:TLL_TLR}
\end{equation}
To calculate the commutator appearing in the current formula we are interested in expressions containing $\check{\mathcal{T}}_{LL}  \circ \check{\mathcal{G}}_{\infty,L} $ and $\check{\mathcal{G}}_{\infty,L} \circ \check{\mathcal{T}}_{LL} $. To find them we start with 
\begin{equation}
\check{\mathcal{T}}_{LL}  =  \check{v}_{LR}  \circ \check{\mathcal{G}}_{\infty,R} \circ \check{v}_{RL}  
\circ \left(\check{1} + \check{\mathcal{G}}_{\infty,L} \circ \check{\mathcal{T}}_{LL} \right)
\end{equation}
and performing 
\begin{equation}
\check{\mathcal{T}}_{LL}  \circ \left(\check{1} + \check{\mathcal{G}}_{\infty,L} \circ \check{\mathcal{T}}_{LL} \right)^{-1} =  \check{v}_{LR}  \circ \check{\mathcal{G}}_{\infty,R} \circ \check{v}_{RL}  
\end{equation}
we are now able to make use of the very important relation $a \circ \left( 1 + b \circ a \right)^{-1} =  \left( 1 + a \circ b \right)^{-1} \circ a$ and we  get
\begin{equation}
\left(\check{1} + \check{\mathcal{T}}_{LL}  \circ \check{\mathcal{G}}_{\infty,L} \right)^{-1} \circ \check{\mathcal{T}}_{LL}  =  \check{v}_{LR}  \circ \check{\mathcal{G}}_{\infty,R} \circ \check{v}_{RL}  
\end{equation}
or
\begin{equation}
\check{\mathcal{T}}_{LL}  = \left(\check{1} + \check{\mathcal{T}}_{LL}  \circ \check{\mathcal{G}}_{\infty,L} \right) \circ \check{v}_{LR}  \circ \check{\mathcal{G}}_{\infty,R} \circ \check{v}_{RL}  
\end{equation}
Now we are able to replace the left hand side of Eq.~(\ref{eq:TLL_TLR}) with the expression above and after removing $\check{\mathcal{G}}_{\infty,R} \circ \check{v}_{RL} $ we find 
\begin{equation}
\check{\mathcal{T}}_{LR} = \left(\check{1} + \check{\mathcal{T}}_{LL}  \circ \check{\mathcal{G}}_{\infty,L} \right) \circ \check{v}_{LR}
\label{eq:connect1}
\end{equation}
With basically the same idea we can find a relation between $\check{\mathcal{T}}_{LL}$ and $\check{\mathcal{T}}_{RL}$. We start with 
\begin{equation}
\check{\mathcal{T}}_{RL}  =  \left(\check{1} - \check{v}_{RL}  \circ \check{\mathcal{G}}_{\infty,L} \circ \check{v}_{LR}  \circ  \check{\mathcal{G}}_{\infty,R} \right)^{-1} \circ \check{v}_{RL} 
\end{equation}
Multiplication of $\check{\mathcal{G}}_{\infty,L}$ from the right and using again the relation $a \circ \left( 1 + b \circ a \right)^{-1} =  \left( 1 + a \circ b \right)^{-1} \circ a$ leads to
\begin{eqnarray}
& & \check{\mathcal{T}}_{RL}  \circ \check{\mathcal{G}}_{\infty,L} = \check{v}_{RL}  \circ \check{\mathcal{G}}_{\infty,L} \nonumber \\
& & \circ  \left(\check{1} -  \check{v}_{LR}  \circ  \check{\mathcal{G}}_{\infty,R} \circ \check{v}_{RL}  \circ \check{\mathcal{G}}_{\infty,L} \right)^{-1}
\end{eqnarray}
Another multplication from the right, now with $\check{v}_{LR}$, let us compare the right hand side of the equation with the definition of $\check{\mathcal{T}}_{LR}$ and we find
\begin{eqnarray}
\check{\mathcal{T}}_{RL}  \circ \check{\mathcal{G}}_{\infty,L} \circ \check{v}_{LR} & = & \check{v}_{RL}  \circ \check{\mathcal{G}}_{\infty,L} \circ \check{\mathcal{T}}_{LR} \label{eq:connect3} \\
\check{\mathcal{T}}_{LR}  \circ \check{\mathcal{G}}_{\infty,R} \circ \check{v}_{RL} & = & \check{v}_{LR}  \circ \check{\mathcal{G}}_{\infty,R} \circ \check{\mathcal{T}}_{RL} \label{eq:connect4}
\end{eqnarray}
where the second equation has been found by interchanging $L$ and $R$. With this connection between $\check{\mathcal{T}}_{LR}$ and $\check{\mathcal{T}}_{RL}$ we can finally find another useful relation by substituting the expression for $\check{\mathcal{T}}_{LR}$ from Eq.~(\ref{eq:connect1}) on the right hand side and removing the $\check{\mathcal{G}}_{\infty,L} \circ \check{v}_{LR}$ factor
\begin{equation}
\check{\mathcal{T}}_{RL} = \check{v}_{RL} \circ \left(\check{1} + \check{\mathcal{G}}_{\infty,L} \circ \check{\mathcal{T}}_{LL} \right) 
\label{eq:connect2}
\end{equation} 

\subsection{Calculation of the spectral current density}
\label{sub:app13}
The spectral current density can be calculated by Eq.~(\ref{eq:spectral_current}). In the following we will use the abbreviation: 
\begin{equation}
K_\epsilon^K = \left[  \check{\mathcal{T}}_{LL}, \check{\mathcal{G}}_{\infty,L} \right]_\circ^K
\end{equation}
If we evaluate the commutator and  perform the matrix multiplications in Keldysh space we get
\begin{eqnarray}
K_\epsilon^K  & = & \hat{\mathcal{T}}_{LL}^R \circ \hat{\mathcal{G}}_{\infty,L}^K + \hat{\mathcal{T}}_{LL}^K \circ \hat{\mathcal{G}}_{\infty,L}^A  \\
& & -  \left( \hat{\mathcal{G}}_{\infty,L}^R  \circ \hat{\mathcal{T}}_{LL}^K + \hat{\mathcal{G}}_{\infty,L}^K \circ \hat{\mathcal{T}}_{LL}^A \right) \nonumber
\end{eqnarray}
In the next step we can replace the Keldysh components of the transfer-matrices by its corresponding retarded and advanced components employing the relation of appendix \ref{sub:app11}:
\begin{equation}
\hat{\mathcal{T}}_{LL}^K = \hat{\mathcal{T}}_{LL}^R \circ \hat{\mathcal{G}}_{\infty,L}^K \circ \hat{\mathcal{T}}_{LL}^A
+ \hat{\mathcal{T}}_{LR}^R \circ \hat{\mathcal{G}}_{\infty,R}^K \circ \hat{\mathcal{T}}_{RL}^A
\end{equation}
Now the commutator takes the form 
\begin{eqnarray}
K_\epsilon^K & = & \hat{\mathcal{T}}_{LL}^R \circ \hat{\mathcal{G}}_{\infty,L}^K \circ \left( \hat{1} + \hat{\mathcal{T}}_{LL}^A \circ \hat{\mathcal{G}}_{\infty,L}^A \right) \nonumber \\
& & - \left( \hat{1} + \hat{\mathcal{G}}_{\infty,L}^R \circ \hat{\mathcal{T}}_{LL}^R  \right) \circ \hat{\mathcal{G}}_{\infty,L}^K \circ \hat{\mathcal{T}}_{LL}^A \nonumber \\
& & +\hat{\mathcal{T}}_{LR}^R \circ  \hat{\mathcal{G}}_{\infty,R}^K \circ \hat{\mathcal{T}}_{RL}^A \circ  \hat{\mathcal{G}}_{\infty,L}^A  \\
& &  - \hat{\mathcal{G}}_{\infty,L}^R \circ \hat{\mathcal{T}}_{LR}^R \circ  \hat{\mathcal{G}}_{\infty,R}^K \circ \hat{\mathcal{T}}_{RL}^A 
\nonumber
\end{eqnarray}
In the last step we will replace the reflection components of the transfer matrix by the corresponding transmission components using the relations of appendix~\ref{sub:app12}. We start by substituting the brackets using Eqs.~(\ref{eq:connect1}) and (\ref{eq:connect2})
\begin{equation}
\hat{\mathcal{T}}_{LR}^A \circ \hat{v}_{LR}^{-1} = \left(\hat{1} + \hat{\mathcal{T}}_{LL}^A  \circ \hat{\mathcal{G}}_{\infty,L}^A \right)
\end{equation}
and
\begin{equation}
\hat{v}_{RL}^{-1} \circ \hat{\mathcal{T}}_{RL}^R  = \left(\hat{1} + \hat{\mathcal{G}}_{\infty,L}^R \circ \hat{\mathcal{T}}_{LL}^R  \right)
\end{equation}
The remaining two reflection components can be eliminated employing Eq.~(\ref{eq:TLL_TLR}) and Eq.~(\ref{eq:connect4}) leading to 
\begin{equation}
\hat{\mathcal{T}}_{LL}^A = \hat{\mathcal{T}}_{LR}^A \circ \hat{\mathcal{G}}_{\infty,R}^A \circ \hat{v}_{RL}, \;
\hat{\mathcal{T}}_{LL}^R = \hat{v}_{LR} \circ \hat{\mathcal{G}}_{\infty,R}^R \circ \hat{\mathcal{T}}_{RL}^R  
\end{equation}
and we can finally write down the commutator corresponding to the spectral current density as
\begin{eqnarray}
K_\epsilon^K & = &\hat{v}_{LR} \circ \hat{\mathcal{G}}_{\infty,R}^R \circ \hat{\mathcal{T}}_{RL}^R \circ \hat{\mathcal{G}}_{\infty,L}^K \circ \hat{\mathcal{T}}_{LR}^A \circ \hat{v}_{LR}^{-1} \nonumber \\
& & - \hat{v}_{RL}^{-1} \circ \hat{\mathcal{T}}_{RL}^R \circ \hat{\mathcal{G}}_{\infty,L}^K \circ \hat{\mathcal{T}}_{LR}^A \circ \hat{\mathcal{G}}_{\infty,R}^A \circ \hat{v}_{RL} \nonumber \\
& & +\hat{\mathcal{T}}_{LR}^R \circ  \hat{\mathcal{G}}_{\infty,R}^K \circ \hat{\mathcal{T}}_{RL}^A \circ  \hat{\mathcal{G}}_{\infty,L}^A  \\
& &  - \hat{\mathcal{G}}_{\infty,L}^R \circ \hat{\mathcal{T}}_{LR}^R \circ  \hat{\mathcal{G}}_{\infty,R}^K \circ \hat{\mathcal{T}}_{RL}^A 
\nonumber
\end{eqnarray}
Performing the quasiclassical limit this expression defines the spectral current as it is given by Eq.~(\ref{eq:neqcurrent}).

\end{document}